\documentclass{article}

\usepackage{arxiv}

\usepackage[utf8]{inputenc} % allow utf-8 input
\usepackage[T1]{fontenc}    % use 8-bit T1 fonts
\usepackage{hyperref}       % hyperlinks
\usepackage{url}            % simple URL typesetting
\usepackage{booktabs}       % professional-quality tables
\usepackage{amsfonts}       % blackboard math symbols
\usepackage{nicefrac}       % compact symbols for 1/2, etc.
\usepackage{microtype}      % microtypography
\usepackage{lipsum}
\usepackage{graphicx}

\usepackage{mathtools}
\usepackage{hyperref}

\newcommand{\R}{\mathbb{R}}
\DeclareMathOperator{\inter}{int}
\DeclarePairedDelimiter\floor{\lfloor}{\rfloor}
\DeclarePairedDelimiter\inner{\langle}{\rangle}
\begin{document}
This work has been submitted to the IEEE for possible publication. Copyright may be transferred without notice, after which this version may no longer be accessible.
\title{Conductivity and Electrode Placement Variability: Evaluating the Cramer-Rao Lower Bound in 2D Electrical Impedance Tomography}
\author{
 Georgii Kormakov \\
  Department of Computational Data Science and Engineering\\
  Skoltech\\
  Moscow, Russia \\
  \texttt{Georgiy.Kormakov@skoltech.ru} \\
  %% examples of more authors
   \And
 Nikolay Koshev\\
   Vladimir Zelman Center for Neurobiology and Brain Rehabilitation\\
  Skoltech\\
  Moscow, Russia \\
  \texttt{N.Koshev@skoltech.ru} \\
  %% \AND
  %% Coauthor \\
  %% Affiliation \\
  %% Address \\
  %% \texttt{email} \\
  %% \And
  %% Coauthor \\
  %% Affiliation \\
  %% Address \\
  %% \texttt{email} \\
  %% \And
  %% Coauthor \\
  %% Affiliation \\
  %% Address \\
  %% \texttt{email} \\
  \thanks{This work was supported in part by the Russian Science Foundation under grant 22-71-10120.}
}
\maketitle

\begin{abstract}
Electrical Impedance Tomography (EIT) is a non-invasive imaging modality that has earned significant attention for its potential in real-time monitoring of various physiological parameters. Despite its promise, the sensitivity and resolution of EIT remain areas of active research. This article presents a new way to analyse the sensitivity of EIT systems using the Cramer-Rao Lower Bound (CRLB). This statistical tool provides a theoretical lower bound on the variance of unbiased estimators. To simulate the EIT full cycle, we compared two forward models: standard Dirichlet-to-Neumann (DtN) and restricted DtN. By taking the CRLB over modelled with Finite Elements Method (FEM) 2D circular area with various conductivity patterns, we showed the influence of electrode configuration on each basic FEM element sensitivity. Also, the sensitivity on small perturbations for different conductivity patterns was offered. This article aims to bridge the gap between the research of widely used hardware implementations standard forward DtN model and theoretically approved restricted DtN. We conclude with a discussion on the motivation behind this research and the objectives our work introduces for the inverse restricted DtN model in practice.
\end{abstract}

\section{Introduction}
Electrical Impedance Tomography (EIT) is a non-invasive, radiation-free medical imaging technique that allows for real-time dynamic assessment of tissue conductivity within the human body. As a modern example of application, EIT was used for monitoring mechanical ventilation, which is important for treating and recovering COVID-19~\cite{eit_ventilation2021}. Unlike traditional imaging methods such as computed tomography (CT) or ultrasound, EIT provides bedside monitoring and offers several unique features~\cite{eit_applications2021}:
\begin{itemize}
    \item Low Cost: EIT systems are relatively inexpensive compared to other imaging modalities
    \item Portable: EIT systems are affordable and portable.
    \item High Temporal Resolution: EIT provides real-time imaging, which is valuable for monitoring physiological changes. It allows EIT to complement ultrasound by offering functional information.
    \item Safety: EIT does not involve ionizing radiation, making it safe for patients.
\end{itemize}

From the opposite point of view, EIT offers many approaches to solve its low spatial resolution~(compared to techniques like MRI or CT) and eliminates high dependence on precise electrode placement and boundary modelling using robust methods\cite{robust_eit2022}. Our goal is also to present a different method to analyse the robustness of the EIT forward model.

% Current State of EIT
EIT has made significant progress in recent years. When contrasts in the tissue's electrical characteristics are produced by the anatomical or physiological phenomena of interest, EIT is helpful. Anatomical contrasts include, for example, the changed electrical impedance spectra of ischemia~\cite{b15} and malignant~\cite{b14} tissues. Functional contrasts are produced when conductively contrasting fluids or gases move in response to changes in tissue conductivity, as occurs during breathing, blood flow, digestion, or neurological activity.

\begin{figure}[htbp]
\centerline{\includegraphics[width=0.5\textwidth]{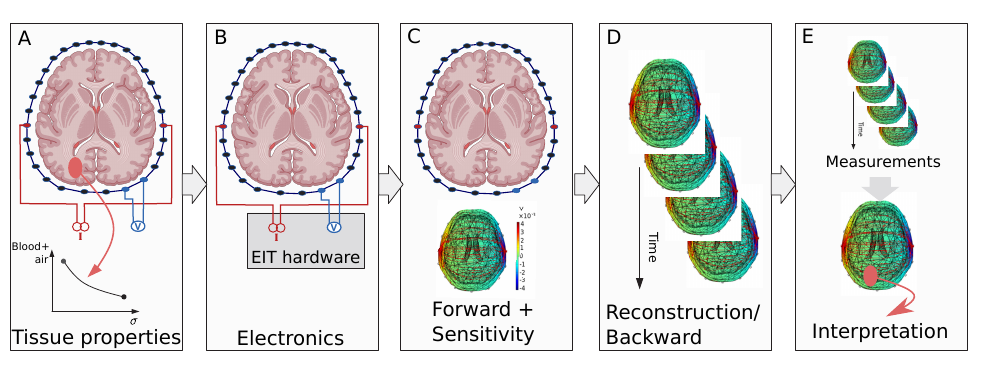}}
\caption{Full brain EIT cycle}
\label{fig:full_eit_cycle}
\end{figure}

The general way to receive the EIT image is described in Fig.~\ref{fig:full_eit_cycle}. 
Because of the discussed features of brain tissues, the EIT electrodes can receive the meaningful forward model image (step C). After that, we must somehow distinguish the changes in the brain. So, the EIT device measures signals on specific frequencies on the predefined period (step D). After all, a researcher can evaluate any measurements from absolute values or any feature type.

% Reformulate simpler!
For a given conductivity distribution within the body, the prediction of measured potentials' values, $U$, is the \textit{forward problem},
which also includes analysis of the sensitivity of $U$ to changes in conductivity~\cite{b11}. Using $N_e$ electrodes, an EIT system calculates a frame of $N_M$ data measurements, $U$. Using pair-drive and avoiding measurements on the driven electrodes,
$N_M = N_E(N_E - 3)$. The maximum number of independent measurements possible on $N_E$ electrodes is $\frac{1}{2}N_E(N_E - 1)$, due to reciprocity (i.e. the sensitivity is unchanged if drive and measurements are interchanged). In the EIT literature, the reciprocity principle is commonly cited as~\cite{b12}, although it was known much earlier~\cite{b13}. Measurements, $U$,
can have complex values indicating the in and out-of-phase components, although it is common to only consider the in-phase component, which dominates at low frequencies.

Advances in instrumentation and image reconstruction methods have expanded EIT applications. However, challenges remain, including robustness and spatial resolution.
In this paper, we studied the sensitivity of two types of forward models, which we'll describe in the first section. The inverse EIT problem study is not discussed in this paper, but it will be based on the restricted model with several theoretical guarantees~\cite{b4}. The article by Klibanov~et~al. also contains several examples of received solutions for this model. Generally, the inverse problem was solved using other models~\cite{Strauss_Khan_2023}~\cite{four_inv_eit_2021}~\cite{Kirsch2021}.

The main goal of the forward problem cycle is to model the object with some mesh and receive the sensitivity of the forward solution to restore the real conductivity of tissues. The first section also describes the common way to analyse the sensitivities, while in the second section, we describe the way how to use the Cramer-Rao Lower Bound (CRLB) estimate for sensitivity analysis of the forward model. CRLB is important not only as a widely used statistical instrument for estimating the variance of noisy data~\cite{b10}. It's also becoming crucial when we want to measure the sensitivity of a mathematical forward model in practice without a bias caused by the numerically collected data. This estimate allows the inverse problem, where we have to find the most accurate conductivity distribution, to show the weak elements of forward modelling. 

In the Numerical experiments section, the 2D configuration for experiments is mentioned and results of sensitivity analysis using basic perturbation idea and CRLB estimate are mentioned.

The last sections are dedicated to challenges and opportunities that the proposed approach and future study propositions could meet.

\section{Modelling}
\subsection{EIT forward models}
Consider predefined conductivity $\sigma (\vec{x}): \R^d \rightarrow \R$ of the area of interest $\Omega \subset \R^d$. Here we used the isotropic version of conductivity, in the anisotropic case the conductivity will be a vector function\footnote{In some cases, the quasi-static approximation could be used (see [chapter 5.2.1, Adler, 2021] with a full complex admittivity $\gamma = \sigma + i\omega \varepsilon$)}. Also, the area should be closed and bounded with the piece-wise smooth border $\partial \Omega$. On this border (or near it), we placed electrodes like in the scheme from Fig.~\ref{fig:full_eit_cycle}.

Let's describe the most popular model for practical goals, where the interior part of the area $\inter \Omega$ has no current sources and the measured voltages are included in boundary conditions only. It relies on the so-called \textit{''traditional'' Dirichlet-to-Neumann (DtN)} data~\cite{b3},~\cite{b4},~\cite{b5}.

\begin{figure}[htbp]
    \centering
    \includegraphics[width=0.4\textwidth]{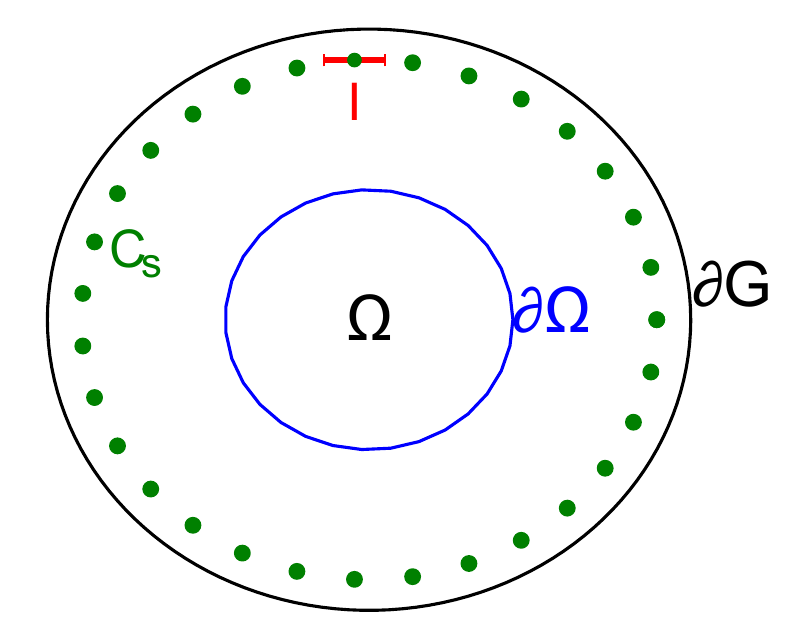}
    \caption{Modelling setup. $\Omega$ ~--- the area of interest (e.g. brain) with the border $\partial \Omega$,  $C_s$ ~--- the set of coordinates of electrodes; $G$ ~--- the extended area with border $\partial G$ for restricted DtN model~\ref{eq:rdtn_main}, $I$ ~--- a small interval including the electrode used mainly for a backward problem solution.}
    \label{fig:forward_sets}
\end{figure}

Because the $\inter \Omega$ has no sources, the main equation inside of the area is simply the Laplace equation (eq.~\ref{eq:dtn_laplace}).
As the boundary conditions, we have the current density $j_{n}(\vec{x})$ measured on $\partial \Omega$ by electrodes (eq.~\ref{eq:dtn_neumann}), where $\vec{n}$ is the outward unit normal to $\partial \Omega$. Also, the Gauss' theorem must be added (eq.~\ref{eq:dtn_gauss}) because eq.~\ref{eq:dtn_neumann} sets the Neumann condition.
\begin{align}
    \nabla \cdot (\sigma (\vec{x}) \nabla U(\vec{x})) &= 0,  \quad \vec{x} \in \Omega  \label{eq:dtn_laplace}\\
    \sigma (\vec{x}) \nabla U(\vec{x})\cdot \vec{n} &= j_{\vec{n}} (\vec{x}),  \quad \vec{x} \in \partial \Omega\label{eq:dtn_neumann} \\
    \int_{\partial \Omega} U(\vec{x}) dS &= 0\label{eq:dtn_gauss}
\end{align}

For the mathematical modelling, the ''restricted'' DtN data can be described~\cite{b4}. The motivation behind the usage of this model is theoretically proved by Klibanov~et~al. advantages for the quality of the inverse solution compared to the classical DtN model~\cite{b4}. This model considers electrodes inside the domain of interest $G$ and sets boundary conditions as zero or elementary potential from sources on its border $\partial G$. With the sources inside the area, the Poisson equation becomes true $\nabla \cdot (\sigma (\vec{x}) \nabla U(\vec{x})) = - \nabla \cdot \vec{J}_S$.
To avoid singularities in the right part, let's change the $\delta$ function of the point source to $\delta$-like $f(x)$ with the following properties
\begin{align*}
    f(x) & \in C^{\infty}(\R^d)\\
    f(x) &\begin{cases}
        = 0, \quad |x| > \varepsilon \\
        \geq 0, \quad |x| \leq \varepsilon \\
        \neq 0, \quad x = 0
    \end{cases}
\end{align*}
Thus, the current density from the right-hand side transforms into the value of a described function of the distance to the source $\vec{x}_s$. In this model, the source is running on the interval $I = \{\vec{x}_s=(x_{1s}, \overline{x})~|~ s \in [0, 1], \text{fixed} \; \overline{x}\in \R^{d-1}\}$ of the straight line\footnote{In this notation, $\vec{x}_s$ is the position of the point source for specific $s \in [0, 1].$}. Original work proposes to take homogeneous Dirichlet condition on the boundary $\partial G$\cite{b4}. For modelling purposes, we propose to use the elementary potential of point source $ g(\vec{x}, s) = 1/(4\pi\varepsilon_0 \|\vec{x} - \vec{x}_s\|)$. Finally, we receive the forward EIT model with ''restricted'' DtN data (eq.~\ref{eq:rdtn_main},~\ref{eq:rdtn_bound}).
\begin{align}
    \nabla \cdot (\sigma (\vec{x}) \nabla U(\vec{x}, s)) &= -f(\vec{x} - \vec{x}_s),  \quad \vec{x} \in G, \;  \vec{x}_s \in \overline{I}\label{eq:rdtn_main}\\
    U(\vec{x}, s)|_{\vec{x}\in\partial G} &= g(\vec{x}, s), \forall \vec{x}_s \in \overline{I}\label{eq:rdtn_bound}
\end{align}
The whole image for the two models is represented in Fig.~\ref{fig:forward_sets}. The last model expects $\Omega$ to be fully inside $G$ (eq.~\ref{eq:sets_cond1}) and the support of the source function $f$ to be outside the $\Omega$ (eq.~\ref{eq:sets_cond2}). Also, the conductivity outside the $\Omega$ is set to 1.
\begin{align}
    \Omega \subset G, \quad \partial \Omega \cap \partial G &=  \emptyset \label{eq:sets_cond1} \\
    I + \mathcal{B}_{\varepsilon}(0) &\subset G \backslash \overline{\Omega}\label{eq:sets_cond2}
\end{align}

\subsection{FEM formulations}
To solve the described equations numerically, Finite Elements Method (FEM) is generally used~\cite{b6},~\cite{b3}. In this method, the variational reformulation is obtained using piece-wise continuously differentiable function $V(\vec{x})$. For the problem with full DtN data (equations~\ref{eq:dtn_laplace} -~\ref{eq:dtn_gauss}), the weak formulation looks following
\begin{align*}
    \int_{\Omega}\sigma (\vec{x}) \nabla U(\vec{x}) \nabla V(\vec{x}) d\vec{V} = \int_{\partial \Omega} j_{\vec{n}} (\vec{x}) V(\vec{x})dS
\end{align*}
Discretising the space of functions $V(\vec{x})$ into the piecewise-linear functions $\phi_i(\vec{x})$ on the basic elements $K \in \mathcal{T}$ of mesh $\mathcal{N} = \{N_i \in \overline{\Omega}, i=\overline{1, M}\}$, the corresponding linear system of equations could be written.
\begin{align}
    \sum\limits_{i, j = 1}^{M} \sum\limits_{(N_i, N_j) \in K} \underbrace{\sigma_K   \int\limits_{K} \nabla \phi_i(\vec{x}) \nabla \phi_j(\vec{x}) d\vec{x}}_{\sigma_K [\mathbf{A_K}]_{ij}} \cdot U(N_i)= \nonumber \\ = \sum\limits_{i=1}^{M} \underbrace{j_{\vec{n}}(N_i) \Phi_i(N_i)}_{\mathbf{I}_i}, \label{eq:weak_dtn}
\end{align}
where $\Phi_i(N_i)$ is the sum over the neighbouring nodes on the border $\partial \Omega$ of basis functions multiplication.
\begin{align*}
     \Phi_i(N_i) = \sum\limits_{j = 1}^{M}  \sum\limits_{\substack{(N_i, N_j) \in K \\ N_i, N_j \in \partial \Omega}}   \int\limits_{K}  \phi_i(\vec{x}) \phi_j(\vec{x})dS
\end{align*}
To rewrite it in the matrix form, a similar construction to the connectivity matrix should be used\cite{b1}. Consider $\mathbf{C} \in \{0, 1\}^{pN\times M}$, where $p$ is the number of points in the basic element of mesh (e.g. 3 for triangulation) and $N$ is the number of basic elements in $\mathcal{T}$. $\mathbf{C}_{ij}$ is equal to 1 if $\floor{i / p}$ is an ordinal number of basic element $K$ and  $N_j \in K$. Also, we can denote $\mathbf{\dot{C}}_k \in \{0, 1\}^{p\times M}$ as the slice of $p$ rows which corresponds to the basic element $k$. Then, the equation~\ref{eq:weak_dtn} transforms into
\begin{align}
    \underbrace{\mathbf{C}^T diag\left(\sigma_1 \mathbf{A_1}, \dots, \sigma_N \mathbf{A_N}\right) \mathbf{C}}_{\mathbf{Y}(\mathbf{\sigma})}\cdot \mathbf{U} = \textbf{I} \label{eq:matrix_dtn}
\end{align}
In the restricted DtN case, we simply receive the equation with the point source function on the right part (without an influence of the Dirichlet condition).
\begin{align*}
    \int_{G}\sigma (\vec{x}) \nabla U(\vec{x}) \nabla V(\vec{x}) d\vec{V} = \int_{G}f(\vec{x} - \vec{x}_s)  V(\vec{x}) d\vec{V} 
\end{align*}
During the discretisation, we should imply the change of variable  $U = g + w$ inside the domain $G$, where $w(\vec{x}, s)|_{\partial G} = 0$. Then, the matrix form receives an additional term $\mathbf{P}$. It corresponds to the admittivity matrix elements out of the area of interest $\Omega$ and doesn't depend on $\sigma$ (because it's equal to 1 outside the $\Omega$). This term is multiplied by the values $\boldsymbol{\Gamma}$ of the function $g$ from the border condition~\ref{eq:rdtn_bound}.
\begin{align}
    \mathbf{Y}(\mathbf{\sigma})\cdot \mathbf{U} = \textbf{F} - \mathbf{P}\cdot \boldsymbol{\Gamma} \label{eq:matrix_rdtn}
\end{align}

So, we can see no difference in the dependence on the conductivity parameter between the two models.
\subsection{Sensitivity}
The dependence of the solution on parameters (called \textit{sensitivity}) in practice approximates true Jacobian $J = \partial U(\vec{x}) \ \partial \sigma$ using several methods - perturbations, admittance matrix differentiation~\cite{b7} and adjoint-field method~\cite{b8}.

The perturbation method will be discussed in the next section. Let's look now at the difference between the sensitivities of described models using admittance matrix differentiation. Because the equivalence of this and adjoint-field ways of measurements was proved (with accuracy to an absolute value)~\cite{b1}. 

Consider one pair of current stimulation $\mathbf{I}_n \in \R^{M\times 1}$, a vector of current values,  and a measurement pattern  $\mathbf{m}_l \in \{0, \pm 1\}^{M\times 1}$, where $[\mathbf{m}_l]_i = \pm 1$ if a node of mesh $N_i$ corresponds to the measuring electrodes: one electrode is positive, another one is negative, and $[\mathbf{m}_l]_i =0$ for non-measuring electrodes. 

Let's receive the solution for the measuring pair from eq.~\ref{eq:matrix_dtn}. It's equal to $\mathbf{U}_n = \mathbf{Y}(\sigma)^{-1} \mathbf{I}_n$ and measured potential is $u_{ln} = \mathbf{m}_l^T \mathbf{U}_n = \mathbf{m}_l^T\mathbf{Y}(\sigma)^{-1} \mathbf{I}_n$. So, to measure the sensitivity to conductivity change in the specific basic element $K$ with ordinal number $k$, the following expression is evaluated:
\begin{align}
    \mathbf{J}_{ln, k} &= \frac{\partial}{\partial \sigma_k}\mathbf{m}_l^T\mathbf{Y}(\sigma)^{-1} \mathbf{I}_n = \mathbf{m}_l^T \frac{\partial}{\partial \sigma_k} \left(\mathbf{Y}(\sigma)^{-1}\right) \mathbf{I}_n = \nonumber \\
    &= - \mathbf{m}_l^T \mathbf{Y}(\sigma)^{-1} \frac{\partial}{\partial \sigma_k} \left(\mathbf{Y}(\sigma)\right)\mathbf{Y}(\sigma)^{-1}\mathbf{I}_n = \nonumber \\
    &= - \mathbf{m}_l^T \mathbf{Y}(\sigma)^{-1} \mathbf{\dot{C}}_k^T\mathbf{A}_k \mathbf{\dot{C}}_k\mathbf{Y}(\sigma)^{-1}\mathbf{I}_n = \nonumber \\
    &= - \left(\mathbf{\dot{C}}_k \mathbf{Y}(\sigma)^{-1}\mathbf{m}_l\right)^T \mathbf{A}_k \left(\mathbf{\dot{C}}_k\mathbf{Y}(\sigma)^{-1}\mathbf{I}_n \right) \label{eq:sensitivity_dtn}
\end{align}

For the restricted DtN data, the expression~\ref{eq:sensitivity_dtn} changes to 
\begin{align*}
\mathbf{J}_{ln, k} = - \left(\mathbf{\dot{C}}_k\mathbf{Y}(\sigma)^{-1}\mathbf{m}_l\right)^T \mathbf{A}_k \left(\mathbf{\dot{C}}_k\mathbf{Y}(\sigma)^{-1}(\textbf{F} - \mathbf{X}\cdot \mathbf{G} )_n \right) 
% \label{eq:sensitivity_rdtn}
\end{align*}

Because $G$ tends to $0$ while the border $\partial G$ goes further away from the electrodes and $F$ tends to the $\delta$-function with smaller $\varepsilon$, the sensitivities for the two models are almost the same.

Another approach to calculating the sensitivity is taking the finite difference approximations of the Jacobian formula
\begin{align}
 S(\Delta \sigma) = \frac{\partial U(\vec{x})}{\partial \sigma} \approx \frac{U(\vec{x}, \Delta \sigma) - U(\vec{x}, \sigma)}{\Delta \sigma}\label{eq:sensitivity_fd}
\end{align}

This method is simpler in implementation but requires more computations (evaluated numerical solutions). It also can be used in functional dependency style to reduce the amount of solver's calls~\cite{b9}. 
Using the eq.~\ref{eq:sensitivity_fd}, we analysed the model with restricted DtN data.

\section{Cramer-Rao Estimate}\label{sec:CRLB_part}
In real data, sensitivity estimation meets systematic and external noise. Let's discuss in this article only the noise of normal distribution origin, which is the often assumption for MEG and EEG sensors data~\cite{skidchenko2023yttrium}~\cite{Razorenova_2024}. To understand what is the behaviour of the modelled solution Cramer-Rao statistical estimate could be used~\cite{b10}.

Consider collected pairs of conductivities $\mathbf{\sigma}^{(n)}$ and corresponding voltages from electrodes $\mathbf{V}_n$ using FEM solver on some predefined (and fixed) mesh $\mathcal{N}$. Because we know that our solution should be in form like equations~\ref{eq:matrix_dtn} and~\ref{eq:matrix_rdtn}, we may rely on the assumption that the received dataset $\{(\mathbf{\sigma}^{(1)},  \mathbf{V}_1), \dots, (\mathbf{\sigma}^{(L)}, \mathbf{V}_L)\}$ contains only additive Gaussian noise component.
\begin{align*}
 \mathbf{V}_n = \mathbf{U}_n(\mathcal{N}, \mathbf{\sigma}^{(n)}) + \mathbf{\varepsilon}_n, & \quad \mathbf{\varepsilon}_n \sim \mathcal{N}(\theta, \Sigma)
\end{align*}

With this assumption, the p.d.f. of one sample is simply a multivariate normal distribution
\begin{align*}
 &p(\mathbf{V}_n, \mathbf{\sigma}^{(n)} | \mathcal{N}) = p(\mathbf{V}_n | \mathbf{\sigma}^{(n)} \mathcal{N}) p(\mathbf{\sigma}^{(n)} | \mathcal{N}) =  \\ 
 &= \frac{1}{(2\pi)^{\frac{M}{2}} |\Sigma|^{\frac{1}{2}}} \exp\left(-\frac{1}{2} (\mathbf{V}_n - \mathbf{U}_n)^T\Sigma^{-1}(\mathbf{V}_n - \mathbf{U}_n)\right) 
\end{align*}

We understood from the previous subsection, that solutions of DtN and restricted one are pretty similar. Let's take into analysis only DtN solution~\ref{eq:matrix_dtn}. Also, we assume for now that the noise variance is distributed independently and homogeneously on each node of mesh $\Sigma = \sigma^2 I_{M\times M}$. 

Then, the log-likelihood of observed data is 
\begin{align*}
 &\mathcal {L}(\mathbf{V}_n, \mathbf{\sigma}^{(n)} | \mathcal{N}) =
 \sum_{n=1}^L -\frac{M}{2}\ln(2\pi \sigma^2) - \frac{1}{2\sigma^2} \|\mathbf{V}_n - \mathbf{U}_n\|_2^2
\end{align*}

Using~\ref{eq:matrix_dtn} and the fact that $\mathbf{Y}^{\top} = \mathbf{Y}$ (because it contains block-diagonal matrix and $\mathbf{A}_k^{\top} = \mathbf{A}_k$) 
\begin{align}
 &\frac{\partial}{\partial \mathbf{\sigma}_k} \mathcal {L} =
 - \frac{1}{2\sigma^2} \sum_{n=1}^L \frac{\partial}{\partial \mathbf{\sigma}_k} \|\mathbf{V}_n - \mathbf{Y}^{-1}(\mathbf{\sigma}^{(n)}) \mathbf{I}_n\|_2^2 = \nonumber \\
 &= - \frac{1}{\sigma^2} \sum_{n=1}^L \left\langle\mathbf{Y}^{-1} \frac{\partial \mathbf{Y}}{\partial \mathbf{\sigma}_k} \mathbf{Y}^{-1} \mathbf{I}_n, \mathbf{V}_n - \mathbf{Y}^{-1} \mathbf{I}_n\right\rangle = \nonumber \\
 &= - \frac{1}{\sigma^2} \sum_{n=1}^L\langle\mathbf{Y}^{-1} \underbrace{\mathbf{\dot{C}}_k^T\mathbf{A}_k \mathbf{\dot{C}}_k}_{\mathbf{B}_k \in \R^{M \times M}} \mathbf{Y}^{-1} \mathbf{I}_n, \mathbf{V}_n - \mathbf{Y}^{-1} \mathbf{I}_n\rangle  = \nonumber \\
 &= - \frac{1}{\sigma^2} \sum_{n=1}^L\langle\mathbf{Y}^{-1} \mathbf{B}_k \mathbf{U}_n, \mathbf{V}_n - \mathbf{U}_n\rangle \label{eq:fisher_jacobian}
\end{align}

To receive a second order Fisher-matrix, let's differentiate the second time
\begin{align}
 \frac{\partial}{\partial \mathbf{\sigma}_k \mathbf{\sigma}_s} \mathcal {L} \cdot \sigma^2 &= &-\sum_{n=1}^L\frac{\partial}{\partial \mathbf{\sigma}_s}\inner*{\mathbf{Y}^{-1} \mathbf{B}_k  \mathbf{Y}^{-1} \mathbf{I}_n, \mathbf{V}_n - \mathbf{U}_n} = \nonumber \\
  &= &-\sum_{n=1}^L\inner*{\frac{\partial \mathbf{Y}^{-1}}{\partial \mathbf{\sigma}_s} \mathbf{B}_k  \mathbf{Y}^{-1} \mathbf{I}_n, \mathbf{V}_n - \mathbf{U}_n}  \nonumber +\\
 & &-\inner*{\mathbf{Y}^{-1} \mathbf{B}_k  \frac{\partial \mathbf{Y}^{-1}}{\partial \mathbf{\sigma}_s} \mathbf{I}_n, \mathbf{V}_n - \mathbf{U}_n}  \nonumber +\\
 & &\inner*{\mathbf{Y}^{-1} \mathbf{B}_k  \mathbf{Y}^{-1} \mathbf{I}_n, \frac{\partial \mathbf{Y}^{-1}}{\partial \mathbf{\sigma}_s}\mathbf{I}_n} = \nonumber \\
 &= & \sum_{n=1}^L\inner*{\mathbf{Y}^{-1} \mathbf{B}_s  \mathbf{Y}^{-1} \mathbf{B}_k  \mathbf{Y}^{-1} \mathbf{I}_n, \mathbf{V}_n - \mathbf{U}_n}  \nonumber +\\
 & &\inner*{\mathbf{Y}^{-1} \mathbf{B}_k  \mathbf{Y}^{-1} \mathbf{B}_s  \mathbf{Y}^{-1} \mathbf{I}_n, \mathbf{V}_n - \mathbf{U}_n}  \nonumber +\\
 & &-\inner*{\mathbf{Y}^{-1} \mathbf{B}_k  \mathbf{Y}^{-1} \mathbf{I}_n, \mathbf{Y}^{-1} \mathbf{B}_s  \mathbf{Y}^{-1}\mathbf{I}_n} = \nonumber \\
  &= &  \sum_{n=1}^L\inner*{\mathbf{Y}^{-1} \mathbf{B}_s \mathbf{Y}^{-1} \mathbf{B}_k  \mathbf{U}_n, \mathbf{V}_n - \mathbf{U}_n}  \nonumber +\\
  & & \inner*{ \mathbf{Y}^{-1} \mathbf{B}_k\mathbf{Y}^{-1} \mathbf{B}_s  \mathbf{U}_n, \mathbf{V}_n - \mathbf{U}_n}  \nonumber +\\
 & &-\inner*{ \mathbf{B}_s \mathbf{Y}^{-1} \mathbf{Y}^{-1} \mathbf{B}_k  \mathbf{U}_n,  \mathbf{U}_n}
 % = & \frac{2}{\sigma^2} \inner*{(\mathbf{Y}^{-1} \mathbf{B}_k)^2   \mathbf{U}_n, \mathbf{V}_n - \mathbf{U}_n}   + \nonumber\\
 % - &\frac{1}{\sigma^2}\inner*{\mathbf{Y}^{-1} \mathbf{B}_k  \mathbf{U}_n, \mathbf{Y}^{-1} \mathbf{B}_k  \mathbf{U}_n}
 \label{eq:fisher_matrix}
\end{align}

If we make the matrix of second-order derivatives from eq.~\ref{eq:fisher_matrix} and then average each element over measured data we will receive the Fisher matrix. In other words, the Fisher matrix's elements are mean values of second-order derivatives on sampled data, which are taken on the FEM basis.
\begin{align*}
\mathbf{F}_{ks} = \\
-&\frac{1}{\sigma^2} \sum_{n=1}^L \mathbb{E}_{\mathcal{N}} \left[ \inner*{\mathbf{Y}^{-1} \mathbf{B}_s \mathbf{Y}^{-1} \mathbf{B}_k  \mathbf{U}_n, \mathbf{V}_n - \mathbf{U}_n}\right]  +  \\
-&\frac{1}{\sigma^2} \sum_{n=1}^L \mathbb{E}_{\mathcal{N}} \left[ \inner*{ \mathbf{Y}^{-1} \mathbf{B}_k\mathbf{Y}^{-1} \mathbf{B}_s  \mathbf{U}_n, \mathbf{V}_n - \mathbf{U}_n}\right]  +  \\
&\frac{1}{\sigma^2}\inner*{ \mathbf{B}_s \mathbf{Y}^{-1} \mathbf{Y}^{-1} \mathbf{B}_k  \mathbf{U}_n,  \mathbf{U}_n}  = \\
=-&\frac{1}{\sigma^2} \sum_{n=1}^L \inner*{\mathbf{Y}^{-1} \mathbf{B}_s \mathbf{Y}^{-1} \mathbf{B}_k  \mathbf{U}_n, \underbrace{\mathbb{E}_{\mathcal{N}} \left[ \mathbf{V}_n - \mathbf{U}_n\right]}_{=0}}  +  \\
-&\frac{1}{\sigma^2} \sum_{n=1}^L  \inner*{ \mathbf{Y}^{-1} \mathbf{B}_k\mathbf{Y}^{-1} \mathbf{B}_s  \mathbf{U}_n, \underbrace{\mathbb{E}_{\mathcal{N}} \left[ \mathbf{V}_n - \mathbf{U}_n\right]}_{=0}}  +  \\
&\frac{1}{\sigma^2}\inner*{ \mathbf{B}_s \mathbf{Y}^{-1} \mathbf{Y}^{-1} \mathbf{B}_k  \mathbf{U}_n,  \mathbf{U}_n}  = \\
= &\frac{1}{\sigma^2} \sum_{n=1}^L\inner*{ \mathbf{B}_s \mathbf{Y}^{-1} \mathbf{Y}^{-1} \mathbf{B}_k  \mathbf{U}_n,  \mathbf{U}_n}  
\end{align*} 

To compute the Fisher matrix values more efficiently, the following representation should be used
\begin{align*}
\mathbf{F}_{ks} = 
&\frac{1}{\sigma^2} \sum_{n=1}^L\inner*{\mathbf{\dot{C}}_s^T\mathbf{A}_s \underbrace{\mathbf{\dot{C}}_s \mathbf{Y}^{-1}}_{\mathbf{Y}^{-1}_s \in \R^{p \times M}}  \underbrace{\mathbf{Y}^{-1} \mathbf{\dot{C}}_k^T}_{\mathbf{Y}^{-T}_k} \mathbf{A}_k \mathbf{\dot{C}}_k \mathbf{U}_n,  \mathbf{U}_n} = \\
= &\frac{1}{\sigma^2} \sum_{n=1}^L\inner*{\mathbf{\dot{C}}_s^T\mathbf{A}_s \mathbf{Y}^{-1}_s  \mathbf{Y}^{-T}_k \mathbf{A}_k \mathbf{\dot{C}}_k \mathbf{U}_n,  \mathbf{U}_n} = \\
= &\frac{1}{\sigma^2} \sum_{n=1}^L\inner*{\mathbf{A}_s \mathbf{Y}^{-1}_s  \mathbf{Y}^{-T}_k \mathbf{A}_k \mathbf{\dot{C}}_k \mathbf{U}_n,  \mathbf{\dot{C}}_s \mathbf{U}_n}
 % \label{eq:fisher_simple}  
\end{align*}

Cramer-Rao Lower Bound (CRLB) estimates the variance of any unbiased estimator. Our ''estimator'' is the derivative of log-likelihood (eq.~\ref{eq:fisher_jacobian}). We'll use the fact that this estimator is unbiased in section~\ref{sec:CRLB_part} during the derivation of Fisher matrix elements.
Then, the estimate for the variance is
\begin{align}
    var(\mathbf{\hat{\sigma}}_k) \geq \frac{1}{L} [\mathbf{F}^{-1}]_{kk} \label{eq:CRLB}
\end{align}

\section{Numerical experiments}
Using CRLB estimate from eq.~\ref{eq:CRLB} and basic approximation of sensitivity by finite difference scheme (eq.~\ref{eq:sensitivity_fd}), we tried to validate the restricted DtN case (eq.~\ref{eq:rdtn_main},~\ref{eq:rdtn_bound}).

To evaluate such estimates the 2-d FEM model with $M = 417$ nodes and $N = 800$ basic elements (Delaunay triangulation) was taken on $\Omega = \{\vec{x}|\;~\|\vec{x}\|\leq 1\}, \quad G = \{\vec{x}|\;~\|\vec{x}\|\leq \sqrt{5}\}$ and 16 electrodes placement on $X_s = \{\vec{x}|\;~\|\vec{x}\| = r_s\}$ with specified radius $r_s$. 

Because a full EIT frame is supposed to take several measurements for not only one pair, several steps (intervals between the electrodes which apply current) were taken.
\begin{figure}[htbp]
\centering
\begin{minipage}{\columnwidth}
\centering
    \includegraphics[height=3.1cm,keepaspectratio]{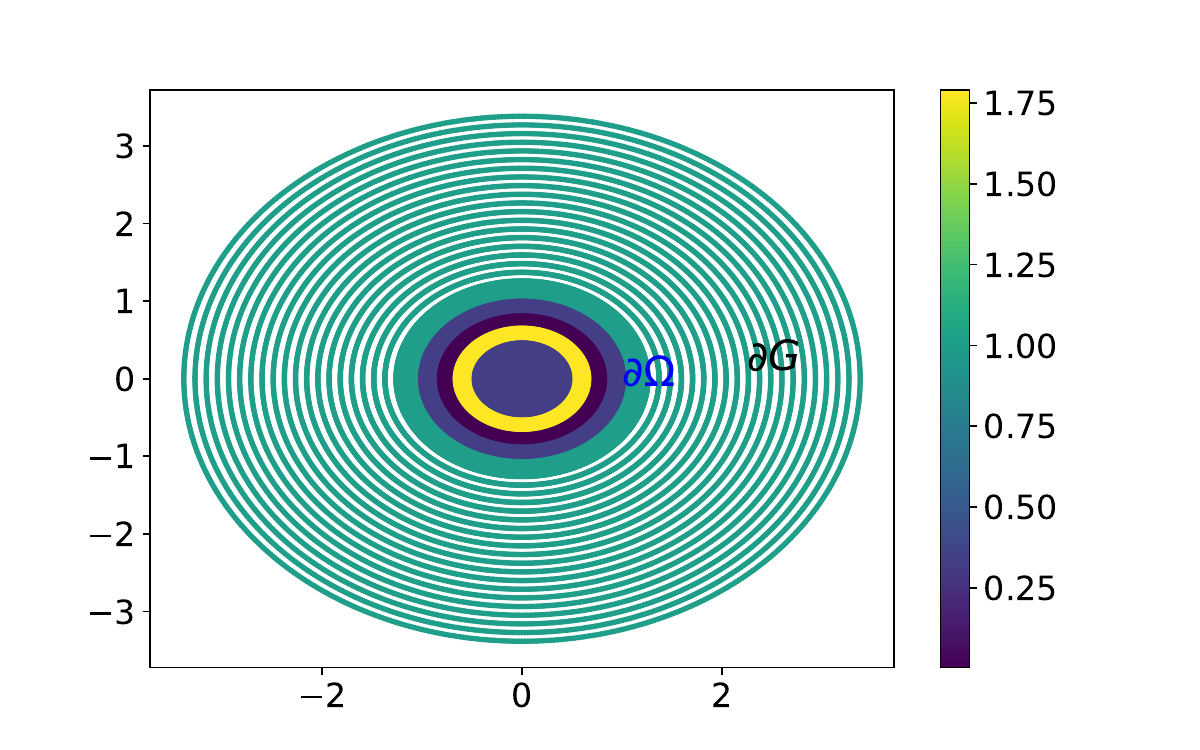}
    \label{subfig:4shell} 
\end{minipage}
\begin{minipage}[t]{\columnwidth}
\centering
    \begin{minipage}{0.49\columnwidth}
    \centering
        \includegraphics[height=2.8cm,keepaspectratio]{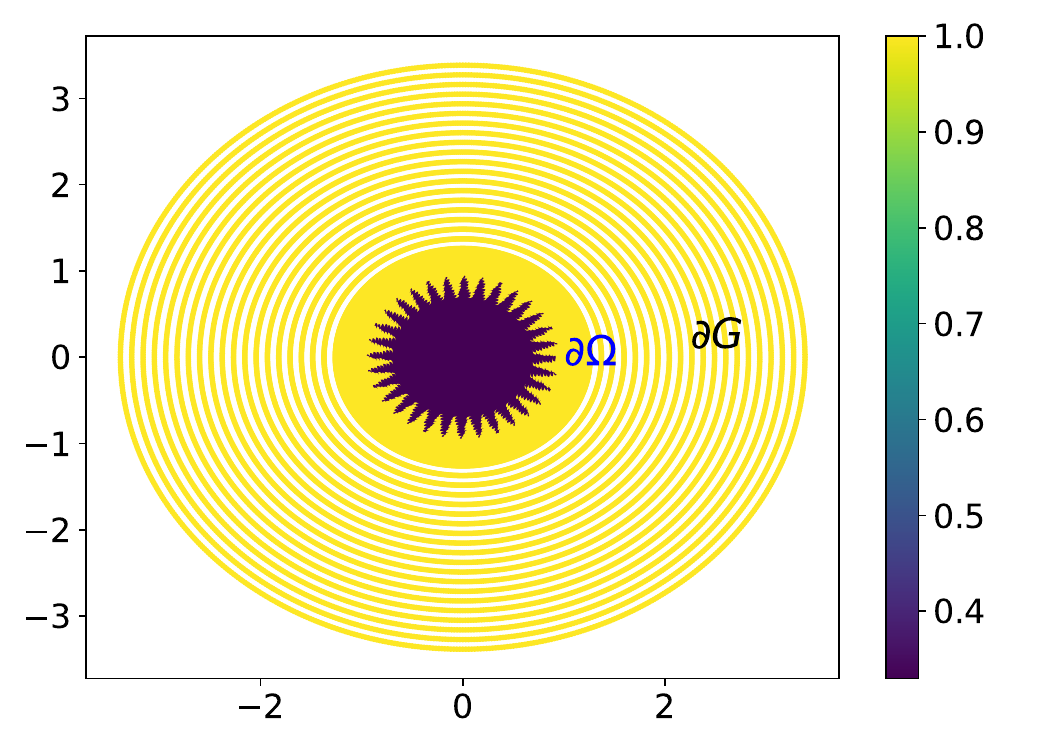}
        \label{subfig:petal}
    \end{minipage}\hfill
    \begin{minipage}{0.49\columnwidth}
    \centering
        \includegraphics[height=2.8cm,keepaspectratio]{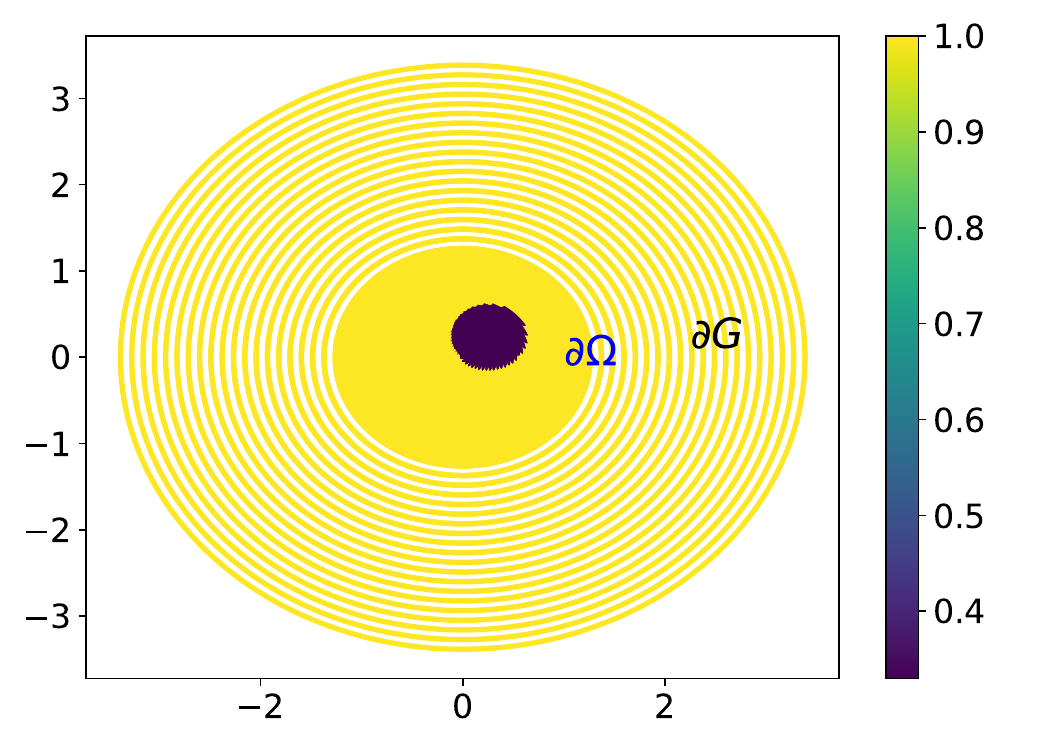}
        \label{subfig:one_source} 
    \end{minipage}
\end{minipage}%
    \caption{Conductivity $\sigma$ patterns used in numerical experiments: 4-shell model (upper), n-petal (left) and 1-circle (right)}
    \label{subfig:conductivity_patterns} 
\end{figure}

\begin{figure}[htbp]
\centering
\begin{minipage}{\columnwidth}
    \centering
    \includegraphics[height=4.3cm,keepaspectratio]{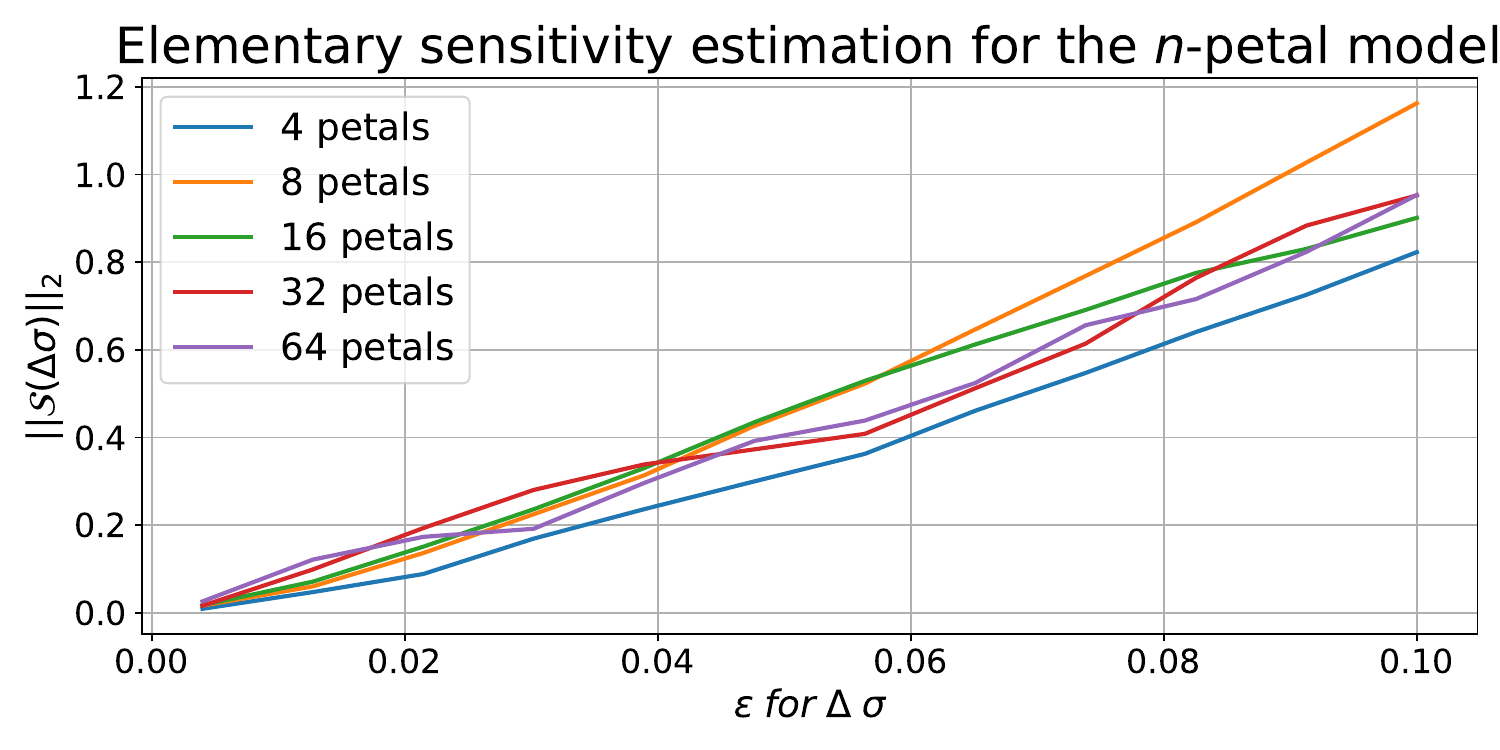}
    \label{subfig:sens_petals} 
\end{minipage}
\begin{minipage}[t]{\columnwidth}
\centering
    \begin{minipage}{0.49\columnwidth}
    \centering
        \includegraphics[height=2.1cm,keepaspectratio]{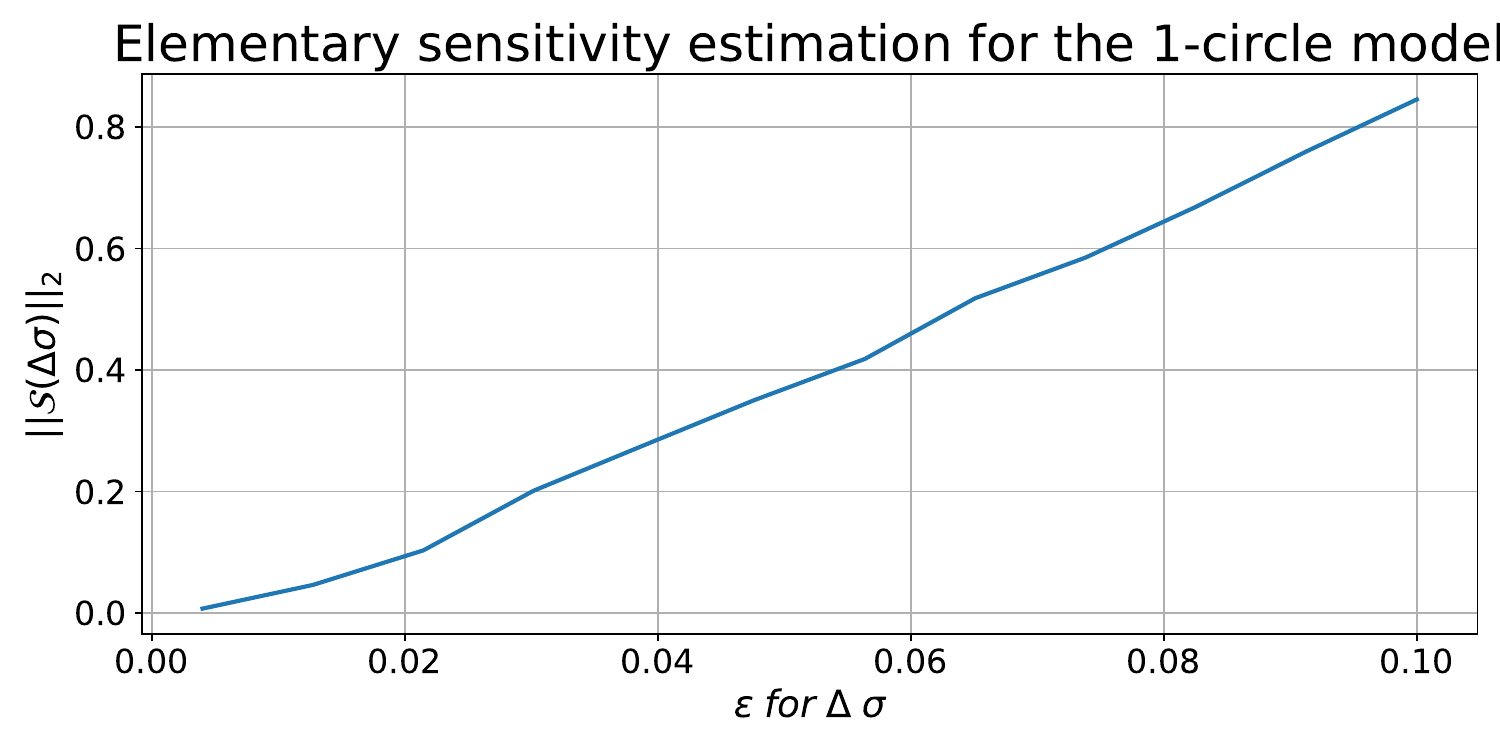}
        \label{subfig:sens_1source} 
    \end{minipage}\hfill
    \begin{minipage}{0.49\columnwidth}
    \centering
        \includegraphics[height=2.1cm,keepaspectratio]{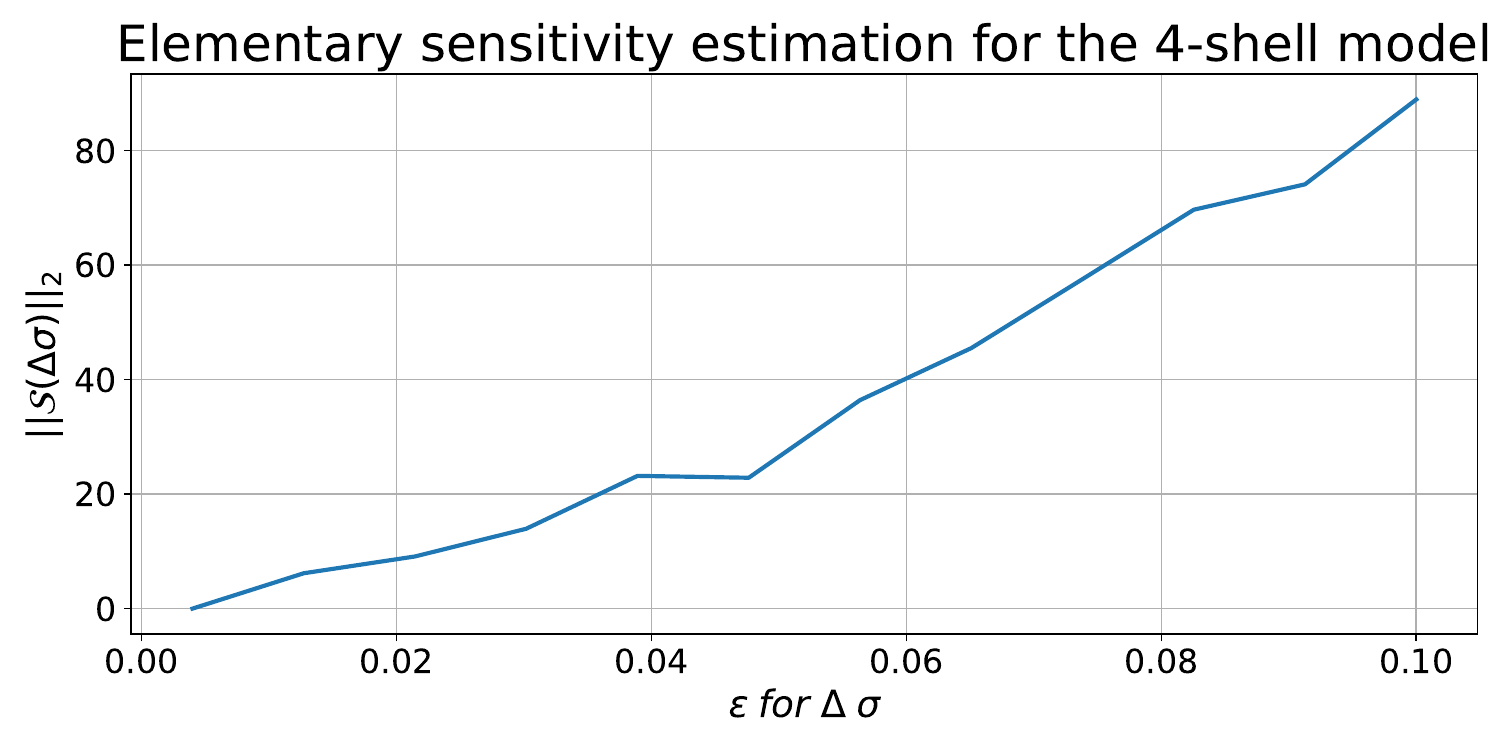}
        \label{subfig:sens_4shell}  
    \end{minipage}
\end{minipage}%
\caption{Elementary sensitivities for $n$-petals (upper), 1-circle (left) and 4-shell (right) models}
\label{fig:elem_sens}
\end{figure}

\begin{figure*}[htbp]
\centering
\includegraphics[width=\linewidth,keepaspectratio]{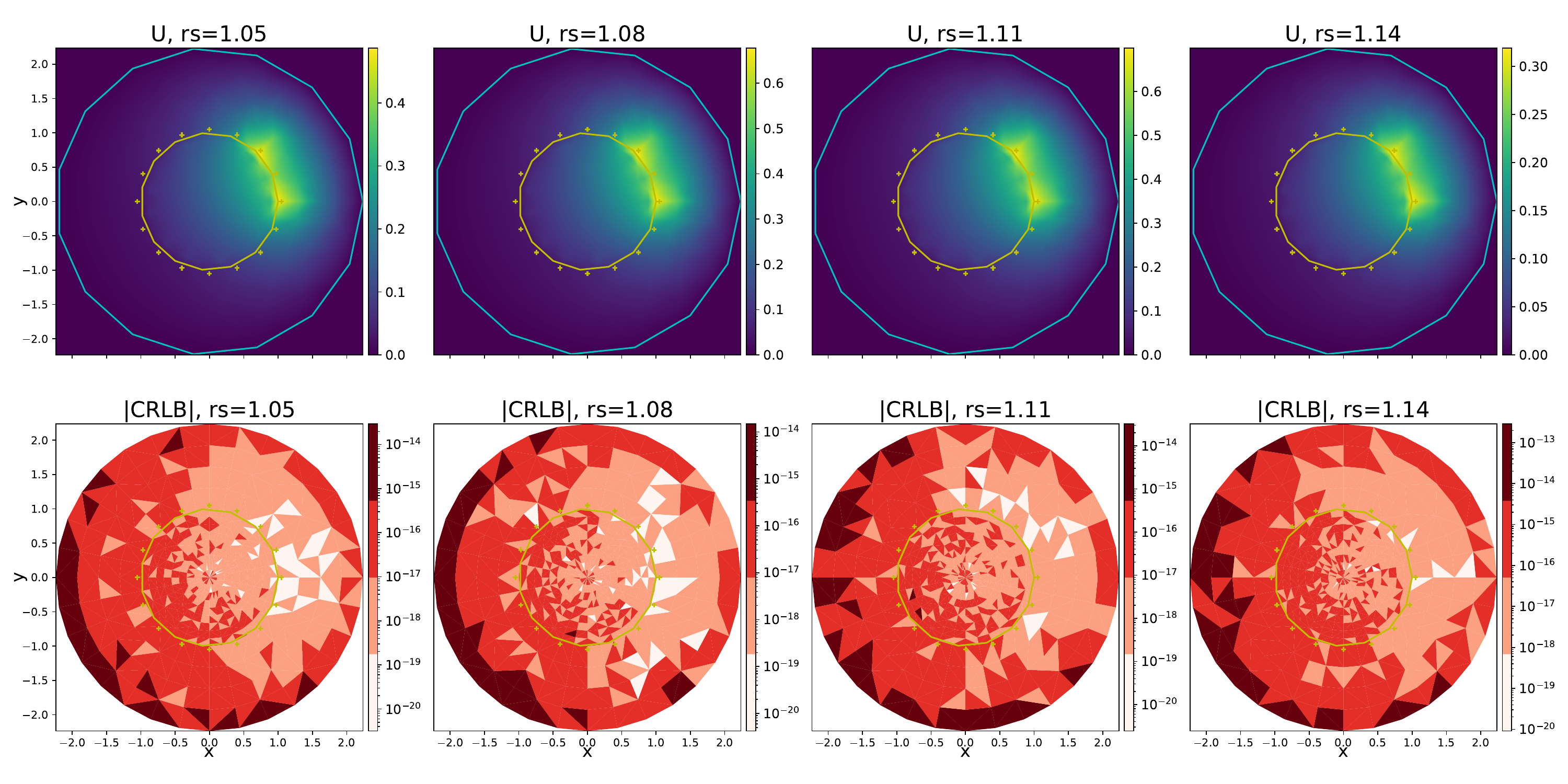} 
    \caption{CRLB for one EIT measurement (with 2 out of 16 applied electrodes with step 1) varying placement radius $r_s$ of electrodes from $1.05$ to $1.14$. Logscale for the CRLB values applied, and they are split into 4 colours to show the main contrasts.}
    \label{fig:crlb_rs_chng}
\end{figure*}

The most useful conductivity distribution for practical research is the so-called 4-shell model~\cite{b16}. The CRLB estimates were received only for this model because it includes more complex changes of conductivity compared with the other two patterns. It has the following 4 layers of conductivities: $(0.3300, 0.0042, 1.7900, 0.3300)$ with the borders on specified relative distances in $cm$ $(0.09, 0.085, 0.08, 0.079)$ and conductivity equals to 1 outside $\Omega$ (fig.~\ref{subfig:conductivity_patterns}, upper).

We also took a 1-circle pattern with one circle of conductivity $0.33$ placed near the electrodes with current (fig.~\ref{subfig:conductivity_patterns}, right) to compare the sensitivity of restricted DtN model with the DtN model results~\cite{b2}.  

Also, the $n$-petal structure proposed to reproduce the patterns of gyrus inside the brain (fig.~\ref{subfig:conductivity_patterns}, left).

\subsection{Basic sensitivity}
To estimate the solution's response to small shifts $\varepsilon$ of the whole pattern of conductivity $\Delta \sigma$ on the grid, we measured finite difference approximation of sensitivity (fig.~\ref{fig:elem_sens}).

The behaviour for each model is the same, but we can see, that even small movements change the solution significantly for the 4-shell model (fig.~\ref{fig:elem_sens}, right). It means that in practice we could see more contrasting results for structures like the 4-shell model (e.g. brain).
Also, changes in the number of petals (fig.~\ref{fig:elem_sens}, upper) don't influence the solution so much. So, the restricted DtN model isn't sensitive to the amount of "gyri-like" patterns.

\subsection{CRLB}
Cramer Rao Lower Bound could help in practice to measure the bounds for estimated conductivity on noisy data. In numerical experiments, such behaviour can only be simulated by pseudo-random noise. So, we decided to receive the CRLB only on one sample of solution and frame solution of EIT (as the mean of many pairs solution).

To understand how the sensitivity changes in practice, it's important to model a typical setting. Thus, we show that after the changing of the source distance to the "skull" the CRLB estimate changes on each element(fig.~\ref{fig:crlb_rs_chng}).

\begin{figure}[htbp]
\centering
\includegraphics[width=0.5\linewidth,keepaspectratio]{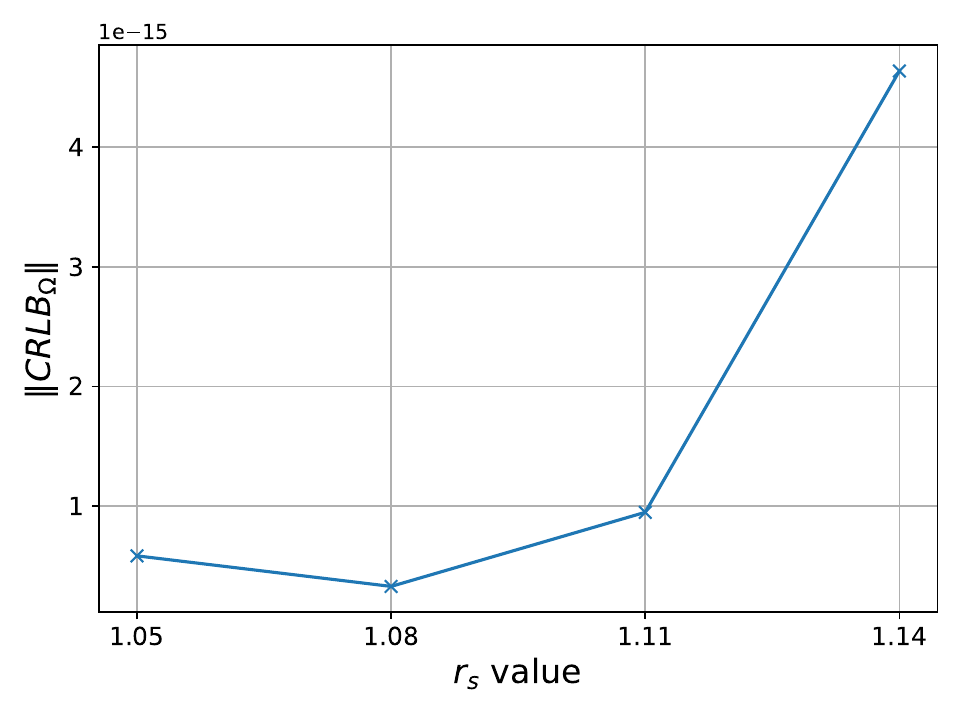} 
    \caption{CRLB inside the $\Omega$ for different values of $r_s$ from fig.~\ref{fig:crlb_rs_chng}}
    \label{fig:4_points_CRLB}
\end{figure}

In addition to artefacts outside the observation area $\Omega$, which are caused by non-uniformly spaced circles in FEM, see points in conductivity patterns (fig.~\ref{subfig:conductivity_patterns}) and the remoteness from the sources, we can see that the estimate for variance in the main area changes insignificantly (about $10^{-15}$), which we measured separately as the norm of CRLB inside the area $\Omega$ (fig.~\ref{fig:4_points_CRLB}). This effect indicates the stability of the model to small fluctuations (while maintaining conductivity).

\begin{figure}[htbp]
\centering
\includegraphics[width=\linewidth,keepaspectratio]{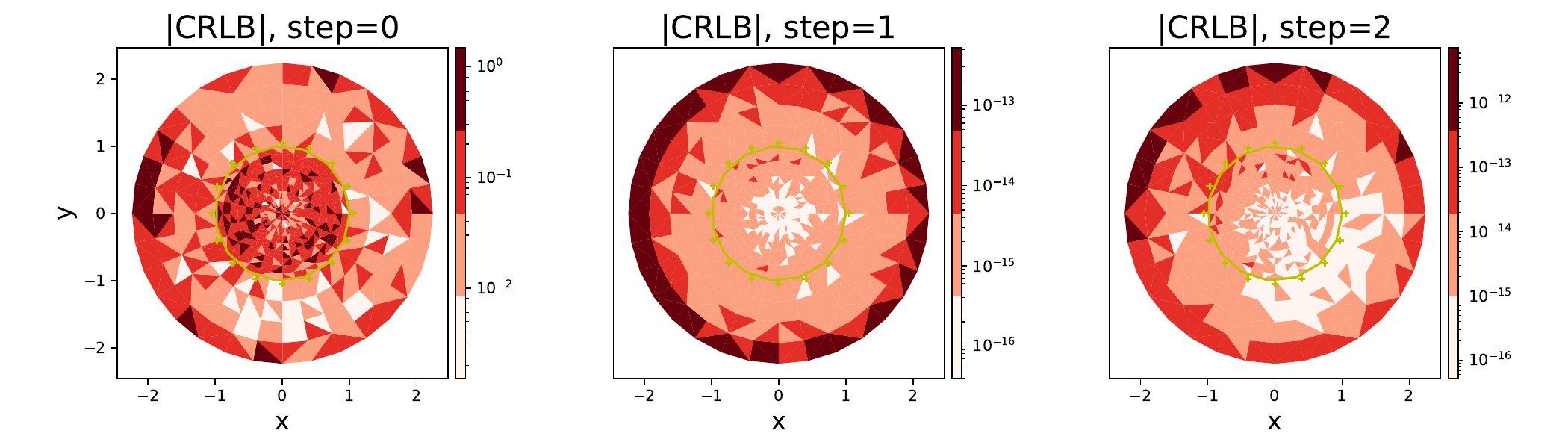} 
    \caption{CRLB for one frame of EIT with different steps between the applied electrodes: 0 (left), 1 (centre) and 2 (right)}
    \label{fig:crlb_step_chng}
\end{figure}

Also, for the EIT technology, the step parameter is crucial~\cite{b3}. So, we decided to model the CRLB for a full EIT solution with 3 steps between electrodes - 0, 1, 2 to see how it can change the sensitivity (fig.~\ref{fig:crlb_step_chng}).

Step 0 (left one on fig.~\ref{fig:crlb_step_chng}) makes the solution unstable, step 1 seems optimal in this case and step 2 shows the non-symmetric distribution of nodes in FEM models. To eliminate such behaviour, a more dense grid should be placed on the modelling area (making mesh-refinement procedure).

\section{Discussion}
The bioimpedance of tissues is relevant because of two main reasons. Firstly, EIT of the brain poses a complicated, but not insuperable, problem because a conductive covering encases the brain, the cerebrospinal fluid, two layers with high resistivities: the pia mater and skull, and then the scalp, which has moderate resistivity. 
The 4-shell model shows some approximation of these natural layers\cite{b16}.
Secondly, there are time-related changes in impedance
in the brain itself, which provide the opportunity for imaging with EIT. 

These time-related changes fall into three main categories: 
\begin{enumerate}
    \item Changes over hours or days related to fluid balance or stroke.
    \item Changes over tens of seconds, due to cell swelling and blood flow, which are relatively large, of the orders of ten to one hundred per cent.
    \item Those due to the opening of ion channels during neuronal activity, which occurs over milliseconds, and are much smaller: $\sim$0.1 – 1\% recorded locally in the brain or nerve and $\sim$2–3 orders of magnitude smaller if recorded on the scalp
\end{enumerate}

To overcome numerical troubles during measurements of these changes, we should analyse the sensitivity of forward models. This article introduces the Cramer-Rao Lower Bound estimation technique for EIT, and several results proposed for the modelled patterns of conductivities, one of which (4-shell) is applied in modern research as an approximation of brain conductivity layers.

For further work, we'll continue to study the restricted DtN model for the EIT, proposed by Klibanov~et~al.~\cite{b4}, which could surely change the quality of EIT imaging. As the next step, we'll model the inverse solution with the technique proposed in~\cite{b4} and analyse it using the results of this research.

\section{Conclusions}
The EIT technology has become more popular in recent years because of the simplicity of modelling and distribution of hardware solutions. But still, we have to explore the modelling procedure to understand the weak places and to find a more robust mathematical model.

In this article, we explored the sensitivity of the theoretically reasonable model of restricted DtN equations. We found that it has the same behaviour on simple examples as the previously studied model.

The method we used is based on a statistical CRLB estimate. It gives the picture of the minimal level of variance of a forward model solution for given conductivity. Thus, using this approach we can show the possibility of a specific forward model usage to approximate a real-life conductivity image. We've visualised the simple numerical results of our method for the restricted DtN data case.

The next step of the research is to study the inverse problem, to use the received CRLB approach for estimating the inverse modelling accuracy and to model more complex patterns (e.g. to estimate the conductivity from MRI and then model a 3-D case).

\bibliographystyle{plain}
\bibliography{literature} 

\end{document}